\documentclass{article}
\usepackage{spconf,amsmath,graphicx, amsfonts}
\usepackage[sort,nocompress]{cite}

\usepackage{booktabs}
\usepackage{adjustbox}
\title{Articulatory Representation Learning Via Joint Factor Analysis and Neural Matrix Factorization}
\name{Jiachen Lian$^{1}$, Alan W Black$^{2}$,  Yijing Lu$^{3}$, Louis Goldstein$^{3}$, Shinji Watanabe$^{2}$, Gopala K. Anumanchipalli$^{1}$}
  
  \address{$^{1}$ UC Berkeley $^{2}$ Carnegie Mellon University $^{3}$ University of Southern California\\
\tt \{jiachenlian, gopala\}@berkeley.edu
  }
\begin{document}
\ninept
\maketitle

\begin{abstract}
Articulatory representation learning is the fundamental research in modeling neural speech production system. Our previous work has established a deep paradigm to decompose the articulatory kinematics data into \textit{gestures}, which explicitly model the phonological and linguistic structure encoded with human speech production mechanism, and corresponding \textit{gestural scores}. We continue with this line of work by raising two concerns: (1) The articulators are entangled together in the original algorithm such that some of the articulators do not leverage effective moving patterns, which limits the interpretability of both gestures and gestural scores; (2) The EMA data is sparsely sampled from articulators, which limits the intelligibility of learned representations. In this work, we propose a novel articulatory representation decomposition algorithm that takes the advantage of \textit{guided factor analysis} to derive the articulatory-specific factors and factor scores. A neural convolutive matrix factorization algorithm is then employed on the factor scores to derive the new gestures and gestural scores. We experiment with the rtMRI corpus that captures the fine-grained vocal tract contours. Both subjective and objective evaluation results suggest that the newly proposed system delivers the articulatory representations that are intelligible, generalizable, efficient and interpretable.
\end{abstract}

\begin{keywords}
Articulatory, Factor Analysis, Gestural Scores, 
\end{keywords}
\section{Introduction}
\label{introduction}

The mainstream research of deep speech representation learning is to develop a human-like neural speech processing system. However, the gap between human and machine intelligence is not that straightforward to be filled in the \textit{post-transformer era}\footnote{We refer to this as the period when transformer architecture is widely adopted in the general areas of artificial intelligence.}~\cite{karita2019comparative, lecun2022autonomous, mohamed2022self_speech, speech_representation}. Inspired by the concept of autonomous human intelligence system~\cite{lecun2022autonomous}, we aim to develop a human speech processing system that is interpretable, efficient and effective. Basically there are two types of such systems. The first one is the neural speech perception system~\cite{1980_speech_perception, perception, perception_production}, which is the theoretical model for most of the current neural networks. The second one is the neural speech production system~\cite{fant1970_production, perception_production} which learns the speech representations in a way that simulates the human speech production process. We focus on the second line. Directly modeling speech production systems from raw speech is fairly difficult and little work has been explored in this direction. To simplify the problem and to make initiative attempts, we model the speech production system and derive the speech representations from the articulatory signal, which we call \textit{articulatory representation learning}. 

Articulatory representation learning is defined over the framework of articulatory phonology~\cite{articulatory-phonology}, which models the relation between phonological representations as a set of discrete units called \textit{gestures}, and the variability in time that derives from variation in the activation of the gestures in real-time: the magnitude of their activation, and the temporal intervals of activation as represented in \textit{gestural scores}. The gestures explicitly capture the moving patterns of different articulators. The gestural scores are the ultimate form of articulatory representations. The desired properties of gestural scores should be as follows: 1) Intelligible. 2) Interpretable. 3) Sparse. The gestures and gestural scores explicitly form a simple yet effective speech production system. Previously, only a few works have attempted to derive the gestures and gestural scores in a data-driven manner.~\cite{spatio-temporal} adopts the convolutive sparse non-negative matrix factorization (CSNMF) to decompose the articulatory data into gestures and gestural scores. Our previous work~\cite{lian_art} proposed an end-to-end neural convolutive matrix factorization (NCMF) paradigm so that both gestures and gestural scores can be automatically learned via neural networks. However, directly applying matrix factorization on the articulatory data is problematic. In the task dynamics model of speech production~\cite{task_dynamical}, each articulator contributes to the production of speech with a certain percentage. These percent contributions should be carefully determined~\cite{sorensen2019task}. In NCMF paradigm~\cite{lian_art}, the gestural scores, which implicitly encode the percent contributions of articulators, are pretty sensitive to the parameter initialization and are randomly determined. This also affects the dynamical patterns of gestures so that some of the gestures do not capture articulator-specific moving patterns, which limits the interpretability of both gestures and gestural scores. 

To alleviate the aforementioned problems, we propose a two-step articulatory decomposition rule. In the first step, we adopt the guided factor analysis algorithm~\cite{guided, sorensen2019task} to extract \textit{factors} and \textit{factor scores} from the articulatory data. The factor characterizes spatial variation in the position and shape of an articulator. The factor scores parameterize how the position and shape of all articulators change over time. The first step ensures that percent contribution of each articulator is within a reasonable range and ensures that the articulator-specific patterns could be captured. In the second step, we perform the NCMF~\cite{lian_art} on the factor scores to to obtain the sparse gestural scores and gestures. As we mentioned, the gestural scores are the ultimate form of articulatory representation. But different from NCMF~\cite{lian_art}, we define the matrix product of gestures and factors as the new \textit{gestures}, which capture fine-grained articulatory moving patterns. Details about our new \textit{gestures} formulation are given in Sec.~\ref{NCMF_sec}. Eq.\ref{gesture_formular}. Combining these two-step decomposition algorithms, we obtain the gestural scores and \textit{gestures} from the articutary data. Another disadvantage of NCMF~\cite{lian_art} is that the MNGU0 data is sparsely sampled from articulators, which limits the intelligibility of learned representations. In this paper,  we experiment with the rtMRI corpus~\cite{rtMRI} that captures the fine-grained vocal tract contours to improve the intelligibility of the learned representations. Both subjective and objective evaluation results suggest that the newly proposed system delivers decent articulatory representations in terms of intelligibility, explainability, efficiency and generalizability. 

\begin{figure*}[!ht]
  \includegraphics[width=17cm,height=7cm]{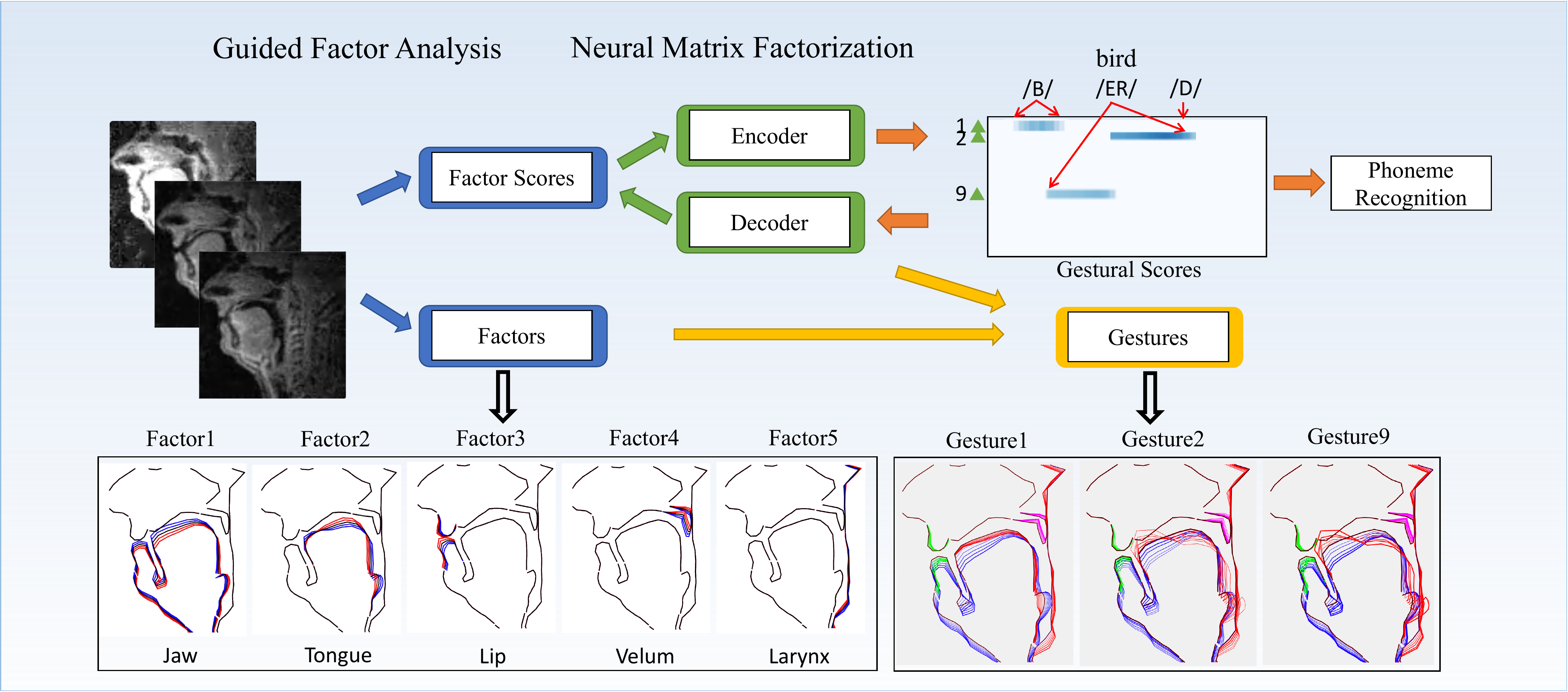}
  \caption{Joint factor analysis and neural matrix factorization paradigm. Each factor characterizes spatial variation in the position and shape of a certain articulator. Gestural scores are the ultimate form of articulatory representations. Gestures are the bases of gestural scores.}
  \label{model}
\end{figure*}

\section{Guided Factor Analysis} 
\subsection{Problem Formulation}\label{problem_formulation}
Given vocal tract video representation $X=[X_1,X_2,...,X_t]\in \mathbb{R}^{2p \times t}$ which is a set of consecutive $t$ frames of vocal tract contours, guided factor analysis~\cite{guided, sorensen2019task} aims to decompose $X$ into articulator-specific \textit{factors} $F=[F_1, F_2, ..., F_q]\in \mathbb{R}^{2p\times q}$ and time domain representation \textit{factor scores} $Y=[Y_1, Y_2, ..., Y_q]\in \mathbb{R}^{t\times q}$, where $2p$ indicates x,y coordinates of contour vertices and $q$ denotes number of factors.  \textit{Guilded factor analysis} can be formulated in Eq.~\ref{factor_analysis_equation}.
\begin{equation} \label{factor_analysis_equation}
X=FY^T=\Sigma_{i=1}^{q}F_i Y_i^T 
\end{equation}

The objective of \textit{guilded factor analysis} is to parameterize the vocal tract representation $X$ into the linear combination~\cite{sorensen2019task} of $q$ factors $F_1, F_2, ..., F_q$ such that each factor characterizes spatial variation in the position and shape of an articulator. The factor scores $Y$ characterizes the temporal variation in the position and shapes of the articulators i.e. factors. In this work, we consider 5 types of articulators: jaw, tongue, lip, velum and larynx. We have one factor for each articulator. Thus, there are 5 factors in total: 
\begin{equation}
    F=[F_{jaw}, F_{tongue}, F_{lip}, F_{velum}, F_{larynx}] 
\end{equation}
\subsection{Factors}
Factors are extracted from the vocal tract representation via eigenvalue decomposition algorithm. In order to obtain the articulator-specific representations, we apply a articulator-specific projection matrix on $X$ such that only the contour points on the current articulator are kept while other entries are set to zero. The projection mechanism can be formulated in Eq.~\ref{projection}, where $art \in \{jaw, tongue, lip, velum, larynx\}$. Denote a binary and diagonal projection matrix as $P_{art} \in \{0,1\}^{2p \times 2p}$. Given an index $i\in [1,p]$, 
the entry $P(2i,2i)$ and $P(2i+1,2i+1)$ which correspond to $x$ and $y$ component of $i$th contour point are 1 if and only if this point is on the current articulator $art$. $P_{art}$ is manually derived via the articulatory segmentation labels of the data~\cite{sorensen2019task}. 

\begin{equation} \label{projection}
    X_{art}=P_{art}^TX
\end{equation}
It is also possible to keep a set of articulators since the motion of some articulators such as \textit{jaw, tongue, lip} co-occur with each other. For example, if we are going to keep both jaw and lip parts, the joint projection will happen: $X_{jaw,lip}=(P_{jaw}^T+P_{lip}^T)X$. In the next step, eigenvalue decomposition is separately applied to extract the articulator-specific factors. Consistent with~\cite{sorensen2019task}, factors are extracted in a two-step manner.

First, we only extract jaw factors which capture the most part of vocal tract motions. Following~\cite{sorensen2019task}, the jaw factors $F_{jaw} \in \mathbb{R}^{2p}$ are defined in Eq.~\ref{jaw_factor}, where $union=\{jaw,tongue,lip\}$. $Q$ and $\Lambda$ denote the eigenvector matrix and diagonal variances matrix for the covariance matrix $X^TX$. For example, $Q_{union}\Lambda_{union}Q_{union}^{-1}=X_{union}^TX_{union}$. Different from~\cite{sorensen2019task}, we do not apply a scaling factor for convariance matrix since it has little effect to the results. Note that $Q_{jaw}$ and $\Lambda_{jaw}$ capture the direction and variance of jaw movement.  Multiplying them with $Q_{union}\Lambda_{union}Q_{union}$ will make the jaw factors capture the joint motion from jaw, lip and tongue articulators. More details can be checked in~\cite{sorensen2019task}. After obtaining jaw factor $F_{jaw}$, the vocal tract data can be recovered via the Eq.~\ref{jaw_recover}, where $F^+$ denotes Moore-Penrose pseudo inverse. 
\begin{equation} \label{jaw_factor}
    F_{jaw}=Q_{union}\Lambda_{union}Q_{union}^{-1}Q_{jaw}\Lambda_{jaw}^{-\frac{1}{2}}
\end{equation}
\vspace{-12pt}
\begin{equation} \label{jaw_recover}
    \hat{X}=XF_{jaw}F_{jaw}^{+}
\end{equation}

Second, we extract the tongue, lip, velum and larynx factors. The motion captured by these factors should be independent of the jaw factors~\cite{sorensen2019task}. In order to remove such dependence, we should remove the jaw factor component from the vocal tract data via the Eq.~\ref{other_data}. In the next step, we still apply the same mask as Eq.~\ref{projection} to obtain the articulatory-specific jaw-free vocal tract contours, which is formulated in Eq.~\ref{projection_jaw_free}. 
\begin{equation} \label{other_data}
    X^{other}=X(I-F_{jaw}F_{jaw}^{+})
\end{equation}
\vspace{-12pt}
\begin{equation} \label{projection_jaw_free}
    X^{other}_{art}=P_{art}^TX^{other}
\end{equation}
Following~\cite{sorensen2019task}, the factors for the a certain articulator $art \in \{tongue, lip, velum, larynx\}$ can be derived via Eq.~\ref{other_factor}, where $Q^{other}_{art}\Lambda^{other}_{art}(Q^{other}_{art})^{-1}=(X^{other}_{art})^TX^{other}_{art}$
\begin{equation} \label{other_factor}
    F_{art}=Q^{other}_{art}\Lambda^{other}_{art}(Q^{other}_{art})^{-1}Q_{art}\Lambda_{art}^{-\frac{1}{2}}
\end{equation}
For all of factors, we just take the first principle component. 
\subsection{Factor scores}
Once the factors for jaw, tongue, lip, velum and larynx articulators have been finalized, the factor scores can be derived via the simple matrix inversion process of Eq.\ref{factor_analysis_equation}. This is shown in Eq.~\ref{factor_score}, where $F^+$ denotes Moore-Penrose pseudo inverse.
\begin{equation} \label{factor_score}
    Y=X^TF^+=X^T[F_{jaw},F_{tongue},F_{lip},F_{velum},F_{larynx}]^{+}
\end{equation}

\section{Neural Convolutive Matrix Factorization} \label{NCMF_sec}
In the factor analysis algorithm, vocal tract data is projected onto articulatory-specific bases called factors and the factor scores are the actual representations that keep the most information of the original data. However, the articulatory representation is also expected to sparse. In this section, we apply the neural convolutive matrix factorization~\cite{lian_art} on factor scores to obtain the sparse representations(gestural scores) and gestures. According to~\cite{lian_art}, the convolutive matrix factorization(reconstruction) is formulated in Eq.~\ref{NCMF}.
\begin{equation} \label{NCMF}
    Y\approx\hat{Y}=\Sigma_{i=0}^{K-1}(\overrightarrow{H}^i)^T\cdot W(i)
\end{equation}
 Now we explain these terms one by one. $\hat{Y}$ denotes the reconstruction of factor scores $Y$. $W\in \mathbb{R}^{K\times D \times q}$ is 1-D convolutional kernel with a kernel size of $K$, input channel size of $q$ and output channel size of $D$. Note that in~\cite{lian_art}, $W$ is called gestures and $D$ is number of gestures. In this work, we still denote $D$ as number of gestures, however, we define the \textit{new gesture} $G\in \mathbb{R}^{K\times D \times 2p}$ in Eq.~\ref{gesture_formular}. $H\in \mathbb{R}^{D\times t}$ is gestural scores, which is the ultimate form of articulatory representation. $\overrightarrow{H}^i$ indicates that $i$ columns of $H$ are shifted to the right. 
\begin{equation} \label{gesture_formular}
    G=WF^T
\end{equation}
According to~\cite{lian_art}, the entire matrix factorization can be implemented via an auto-encoder framework. The \textbf{encoder} takes factor scores $Y$ derived from Eq.~\ref{factor_score} as input and output the gestural scores $H$. The \textbf{decoder} takes the gestural scores $H$ as input to reconstruct the factor scores $\hat{Y}$. The \textbf{encoder} can be any type of neural network $f(\cdot)$ such that $H=max(f(X), 0)$. The \textbf{decoder} is single 1-D convolutional layer with weight matrix $W$ in Eq.~\ref{NCMF}. 

We use almost the same loss objectives as introduced in~\cite{lian_art}. The reconstruction loss is L2 loss. The sparsity loss is computed over all vectors only in time (Eq.~\ref{sparsity loss time dimension}) dimension, where the vector-wise sparsity is defined as $s(H_i)= \frac{\sqrt{t}-\frac{L_1(H_i)}{L_2(H_i)}}{\sqrt{t}-1}$, in consistent with~\cite{sparseness}, where $L_1$ and $L_2$ denote L1 norm and L2 norm. Connectionist Temporal Classification (CTC)~\cite{ctc} loss $L_{CTC}$ is introduced when performing phoneme recognition. $\lambda_{1,2}$ are balancing weights. Different from~\cite{lian_art}, the entropy loss is removed since we experimentally find it less helpful in improving the interpretability of gestural scores.
\begin{equation}\label{sparsity loss time dimension}
S(H)=\frac{1}{D}\Sigma_{i=1}^{D}s(H_i)
\end{equation}
\vspace{-13pt}
\begin{equation} \label{loss resynthesis}
    L=\mathbb{E}_{X}[||Y-\hat{Y}||_2-\lambda_1S(H)+\lambda_2L_{CTC}]
\end{equation}

\section{Experiments} \label{experiments}
\subsection{Dataset and Preprocessing} \label{dataset}
The rtMRI dataset includes eight (four male, four female) speakers of American English~\cite{rtMRI}. All the speaker were asked to read the same visually presented text from a paper card. The speakers produced the sequence of utterances ten times. The vocal tract constrictions were recorded in the real-time MRI videos. The real-time MRI pulse sequence parameters can be checked in~\cite{sorensen2019task}. The video resolution is 83 frames per second. We directly use the segmentation results from~\cite{sorensen2019task} which adopts the span segmentation methods~\cite{segmen}. 5 articulators are involved as mentioned in Sec.~\ref{problem_formulation} and they are: jaw, tongue, lips, velum and larynx. The total number of points for each image is $2p=400$. For phoneme experiments, we also extract the mel-spectrogram(win=1024,hop=256) and WavLM~\cite{chen2022wavlm} base features from the audio waveform. Sampling rate for rtMRI audios is 16k. We use MFA~\cite{mfa} to extract monophones given text-audio pairs and there are in total 72 monophones. 

\subsection{Modules Details}
The entire articulatory representation learning pipeline is shown in Fig.\ref{model}. Guided factor anaysis is done offline and we use the same implementation as~\cite{sorensen2019task}. The encoder takes factor scores and generates the gestural scores, which is fed into decoder to reconstruct the factor scores. We use similar encoder/decoder as~\cite{lian_art}. The encoder consists of 2 convolutional layers and the decoder is a single convolutional layer. Denote $(\cdot, \cdot)$ as $(out\_channels,kernel\_size)$. The configurations for these three layers are $(64, 3), (D,3), (21,2p)$ respectively, where the number of points in a image $2p=400$, number of gestures $D=15$ and window size $K=21$ for convolutive matrix factorization. Note that the weights of the decoder are the "gestures" that is defined in~\cite{lian_art}. In this work, the \textit{gesture} is the multiplication of "gestures" and factors, as shown in Eq.~\ref{gesture_formular}. The phoneme recognizer takes different types of features as input and is optimized via CTC~\cite{ctc} training. For rtMRI data, we simply reshape each frame of image into a hyper-vector so that the entire rtMRI video sequences become a 2D feature. For all the other features, we just keep their original forms. There are two types of phoneme recognizes: base and large. The base model consists of 3-layer multi head attention blocks, where the number of attention heads are 4 and feed-forward layer dimension is 128. The large model consists of 6-layer multi head attention blocks, where the number of attention heads are 8 and feed-forward layer dimension is 256. 

\subsection{Implementation Details}
We use the same way as~\cite{sorensen2019task} to first obtain factors and factor scores. In the next step, we consider two sets of experiments. (1) We perform rtMRI factor scores resynthesis to extract the gestural scores and to derive the gestures. (2) We use various audio features including rtMRI original data, factor scores, gestural scores, mel spectrograms and WavLM~\cite{chen2022wavlm} features to perform phoneme recognition. For gestural scores, we also furthermore fine-tune both encoder and decoder, which is denoted as \textit{Gestural Scores(FT)} in Table.~\ref{Phoneme_Recognition_Results}. For the overall loss function defined in Eq.\ref{loss resynthesis}, we set $\lambda_1=0.5$. By default $\lambda_2=0$ unless we finetune the gestural score where $\lambda_2=0.3$. For all experiments, adam~\cite{kingma2014adam} optimizer is used with an intial learning rate of 0.001 and weight decay of 4e-4. All the experiments are trained for 1k updates with a batch size of 4. The learning rate is decayed every 10 updates with a factor of 0.95. For phoneme recognition experiments, we also vary the training size in a manner that we gradually increase the number of speakers. For each speaker, we have 3 random utterances as test set and the remaining as training set. Beam search with a width of 10 is used for decoding. We evaluate the intelligibility, generaliability and efficency of the gestural scores and the interpretability of both gestural scores and gestures. 

\subsection{Results and discussions}
\subsubsection{Intelligibilty and Generaliability}
The intelligibility of representations is evaluated via phoneme error rate (PER). In Table~\ref{Phoneme_Recognition_Results}, rtMRI achieves a better PER compared to Mel and WavLM because it captures most of the vocal tract information, and the rtMRI audio data is not clean. Since there is information deduction from rtMRI to factor scores and from factor scores to gestural scores, PER also increases correspondingly. Although the PER for gestural scores is the highest compared to the others, it is still a promising number, indicating that the gestural scores are intelligible. WavLM achieves better PER than Mel under all settings, this is in line with~\cite{C-DSVAE, utts}. We also observe that articulatory features (rtMRI, Factor Scores, Gestural Scores) are less sensitive to data and model size. We use "Range" and "Model Variance" to evaluate this. Range is simply the absolute difference between the PER of the largest data size and the PER of the smallest data size. Model variance is the mean square value of PER difference averaged over all speakers. For CTC-Base, we set them as 0 for all features. As shown in the table, both Mel and WavLM have the largest value for "Range" and "Model Variance," indicating that they are pretty sensitive to data and model scale. We conclude that articulatory features are more generalizable representations.
\vspace{-12pt}
\begin{table}[h]
    \centering
    \caption{Phoneme Error Rate(PER) $\%$ and Sparsity.}
    \begin{adjustbox}{width=240pt,center}
    \begin{tabular}{||c c c c c c c c c c c c||} 
     \hline
     \hline
     $\#$Speakers & 1 & 2 & 3 &4 &5 &6 &7 &8 & Range & Model Variance&Sparsity\\ [0.5ex]
     \hline
     \hline
     \multicolumn{12}{||c||}{CTC-Base}\\
     \hline\hline
     rtMRI & 18.4 & 18.3 & 17.5& 16.4 & 16.3& 17.1&15.7&12.1 & 6.3 & / & 0.14\\ 
     Mel & 32.1 & 25.1 & 24.2& 24.1 & 19.7& 16.4&15.3&13.5 & \textbf{18.6}&/&0.33\\ 
     WavLM & 27.2 & 25.2 & 24.1& 20.5 & 18.3& 17.9&14.2&13.0&\textbf{14.2}&/&0.25\\ 
     Factor Scores & 21.2 & 21.7 & 21.4& 19.5 & 18.5& 17.4&16.9&16.2&5&/&0.28\\ 
     Gestural Scores& 22.4 & 22.2 & 21.1& 21.8 & 21.1& 22.2&20.1&19.7&2.3&/&\textbf{0.93}\\
     Gestural Scores(FT)& 22.1 & 21.0 & 21.0& 21.1 & 20.4& 18.7&19.2&17.6&4.5&/&0.80\\
     \hline\hline
     \multicolumn{12}{||c||}{CTC-Large}\\
     \hline\hline
     rtMRI & 16.1 & 16.0 & 15.8& 15.1 & 15.0& 14.8&14.0&11.1&5 & 3.3&0.14\\ 
     Mel & 22.1 & 20.1 & 17.2& 17.1 & 15.7& 13.4&13.3&12.5&\textbf{9.6}&\textbf{31.6}&0.33\\ 
     WavLM & 20.2 & 19.2 & 17.1& 16.5 & 16.3& 14.9&13.2&12.1&\textbf{8.1}&\textbf{20.6}&0.25\\ 
     Factor Scores & 20.2 & 21.2 & 19.4& 18.8 & 18.9& 17.2&16.5&15.2&5&0.88&0.28\\ 
     Gestural Scores& 20.4 & 20.2 & 19.1& 19.8 & 19.2& 18.2&18.3&17.5&2.9&5.5&0.93\\
     Gestural Scores(FT)& 18.5 & 18.1 & 18.0& 17.1 &18.4& 16.7&16.7&16.0&2.5&7.9&0.74\\
     \hline
    \end{tabular}
    \label{Phoneme_Recognition_Results}
    \end{adjustbox}
\end{table}
\vspace{-16pt}
\subsubsection{Sparsity and Efficiency}
Fig.~\ref{model} shows that the derived gestural scores are sparse. We use Eq.\ref{sparsity loss time dimension} to compute the sparsity of the features. As shown in Table.~\ref{Phoneme_Recognition_Results}, the gestural scores have the largest sparsity while reaching the comparable PER values. This indicates that quite a few information in traditional speech features might not be useful in recognition tasks. Fine-tuning will give lower sparsity since we experimentally observe that there is a trade-off between intelligibility and sparsity. We conclude that gestural scores are efficient representations.  

\subsubsection{Interpretability}
\begin{figure}[h]
    \centering
    \includegraphics[width=8.8cm]{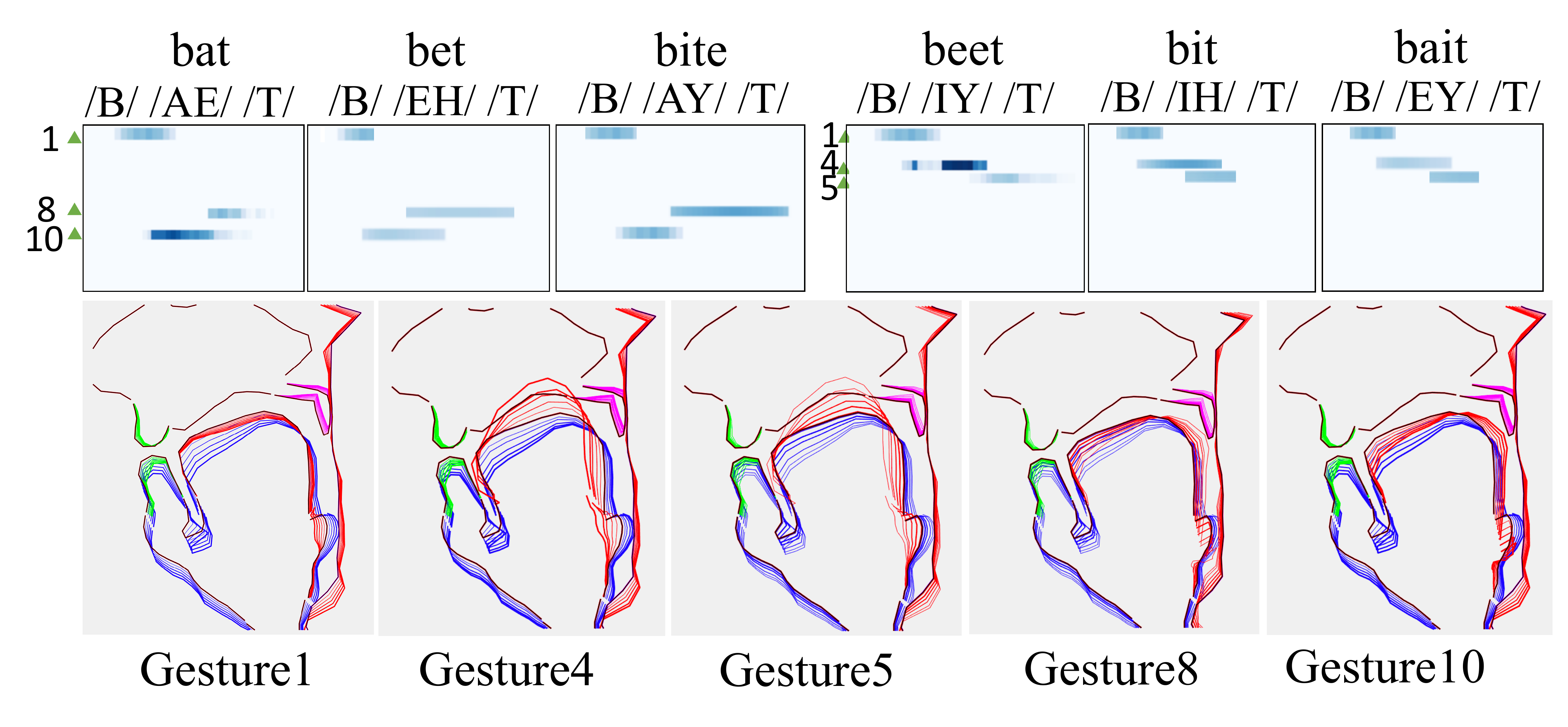}
    \caption{Gestural Scores Visualization.}
    \label{cross_tokens}
\end{figure}

The learned gestural scores are not only intelligible, generalizable and efficient, but also pretty much interpretable. We first look at Fig.~\ref{model}, the gestures 1,2,9 are sequentially activated for the word "bird". Gesture1 corresponds to phone /B/ since the lower lip, jaw, and tongue move down and velum moves up. Gesture 2 corresponds to phone /ER/ since the tongue tip moves up and tongue dorsum moves down. The jaw and lower lip move down and the velum still moves up. Gesture 9 shows opposite moving patterns of gesture 2, indicating the mouth closure. The interpretability of gestural scores is consistent across different tokens. In Fig.~\ref{cross_tokens}, we list two sets of examples. The first set visualizes the gestural score of "bat", "bet" and "bite". We observe that gesture1, gesture8 and gesture10 are activate for all three words. This is totally true since gesture1 corresponds to phone /b/. Regarding the phone /AE/, /EH/ and /AY/, their articulatory patterns are almost the same: lower up moves down, tongue tip and tongue body move down, jaw moves down and velum moves up, as indicated by gesture10. Gesture8 is opposite to gesture10 and it means the closure of mouth. Since gesture10 and gesture8 are able to represent /AE/, /EH/ and /AY/, the only difference among these words is the time interval that is activated for each gesture. For example, in "bat", gesture10 is activate for a longer time interval while in "bite", gesture10 is activate for a shorter interval. The second set visualizes the gestural scores for "beet", "bit" and "bait".  Gesture4 and gesture5 can represent /IY/, /IH/, and /EY/. This is reasonable because when pronouncing these phones, the tongue tip moves down and tongue body moves up. Similarly, the only difference among these words is the gestural scores patterns. In all of the examples, each articulator in the active gestures has reasonable physical meaning while this is not true for the cases in~\cite{lian_art}.  
\vspace{-10pt}
\vspace{-19pt}

\section{Conclusions and Limitations}
Articulatory representation learning is the fundamental methodology for modeling the neural speech production system. We take advantage of guided factor analysis, as well as neural convolutive matrix factorization, to extract the gestural scores and gestures from rtMRI data, a fine-grained vocal tract corpus. The learned articulatory representations, i.e., gestural scores, are intelligible, generalizable, efficient, and interpretable. However, there are still some limitations. First, the presence of noise in rtMRI audio raises concerns regarding the  conclusions about the generalizability. Second, the factor analysis algorithm is independent of neural matrix factorization, which makes the entire representation learning less efficient. Third, the current representations, i.e., gestural scores, have to be derived from articulatory data. However, the neural speech production system should only have speech signals accessible. Deriving an articulatory-free neural speech production system is our future work.
  
\section{Acknowledgement}
This research is supported by the following grants to PI Anumanchipalli --- NSF award 2106928, Rose Hills Foundation and Noyce Foundation.



\bibliographystyle{IEEEbib}
\bibliography{mybib}

\end{document}